# Non-monotonic solutions for colloid-transport of size-distributed particles in porous media

Gabriel Malgaresi, Ben Collins, Paul Alvaro, Pavel Bedrikovetsky

*Australian School of Petroleum, University of Adelaide, Australia*

**Abstract**

Non-monotonic retention profiles (NRP) have been observed in numerous studies of colloidal-nano flows in porous media. For the first time, we explain the phenomenon by distributed particle properties (size, shape, surface charge). We discuss colloidal-nano transport with fines attachment considering stochastically distributed filtration coefficient (particle attachment probability) and the influence of area occupation by particles on the rock surface. The distributed dynamic system allows for exact averaging (upscaling) yielding a novel 3×3 system of equations for total concentrations. Besides the traditional equations of particle mass balance and capture-rate, the novel system contains a third independent equation for kinetics of site occupation during particle attachment. Ten laboratory tests exhibiting NRP have been successfully matched by the upscaled system for binary colloids. Ten laboratory tests with 7-parametric data arrays have been successfully matched by the 5-parameter model, which validates the model. The tuned parameters belong to their common intervals. The laboratory data tuning was significantly simplified by deriving the exact solution of upscaled equations. These results provide valuable insights for understanding the transport mechanisms and environmental impact in colloidal-nano flows exhibiting NRP. Besides, the upscaled system and the analytical model for 1D transport facilitate interpretation of the laboratory coreflood data and allows for the laboratory-based predictions for 3D colloidal-nano transport at the field scale.

**Nomenclature**

$\tilde{B}$ – surface area fraction occupied by a single particle with radius $r$, $L^4$

$b$ – overall concentration of occupied sites on the rock surface

$\check{C}$ – suspended concentration distribution $\check{C}(x,t)$, $L^{-4}$

*c* – total suspended particle concentration, $L^{-3}$

*d* – site occupation function $d(c)$,

*F* – individual suspension function for particles with size *r*, $L^{-5}$

*f* – upscaled suspension function $f(c)$, $L^{-4}$

*h* – filtration function $h(s)$,

*Š* – attached concentration distribution $Š(x,t)$, $L^{-4}$

*s* – total attached particle concentration $s(x,t)$, $L^{-3}$

*t* – time, *T*

*x* – Cartesian coordinate

$f_0$ – fraction of the overall flux

$s_0$ – accessibility factor

*n* – ratio between the filtration coefficients

*Greek letters*

$\lambda$ – filtration coefficient, $L^{-1}$

$\mu$ – viscosity, $ML^{-1}T^{-1}$

$\phi$ – porosity

$\Gamma$ – Henry's adsorbed coefficient

$\varepsilon$ – ratio of occupied areas

## 1. Introduction

Colloidal-nano transport in porous media occurs in numerous chemical engineering and environmental technologies. The examples include but are not limited to industrial filtering, storage of fresh water in underground reservoirs, propagation of viruses and bacteria in subterranean waters, disposal of industrial wastes in aquifers, transport of engineered nanoparticles (NPs), size-exclusion chromatography, and disposal of water produced during oilfield exploitation in aquifers [1-16]. Laboratory-based predictive mathematical modelling determines decision-making in these technologies.

During colloidal, suspension and NP transport in natural porous reservoirs (so-called deep bed filtration), the particles are captured by the rock. The capture mechanisms are size exclusion, attachment, straining, gravitational segregation, bridging, and diffusion into dead-end pores [1-4, 6, 8, 17].

The mathematical model for deep bed filtration adopted in this study consists of mass balance of suspended, captured and adsorbed particles, and capture kinetics [18-22]

$$\frac{\partial}{\partial t}(\phi s_0 c + s + \Gamma c) + U \frac{\partial (f_0 c)}{\partial x} = 0, \tag{1}$$

$$\frac{\partial s}{\partial t} = h(s) f(c) f_0 U. \tag{2}$$

Here $c$ and $s$ are suspended and attached concentrations, respectively, $\phi$ is the porosity, $s_0$ is the accessibility factor, $f_0$ is the fraction of the overall flux moving the accessible pores, $\Gamma$ is Henry's adsorption coefficient, $U$ is the fluid velocity, $h(s)$ is the filtration function, and $f(c)$ is the suspension function. For small $c$, the retention rate is proportional to suspended concentration

$$f(c) = \lambda c, \tag{3}$$

where the filtration coefficient $\lambda$ (probability of particle capture per unit length of the trajectory) is equal to $\lambda = k_d \phi / U$, where $k_d$ is the deposition rate coefficient [19, 23, 24].

Particle capture by a pore is analogous to chemical reaction "pore + particle". Therefore, the capture kinetics (2) is analogous to the active mass law that contains the product of the filtration function that depends on the retained concentration, and the suspension function that depends on the suspended concentration. Like in the active mass law, the functions are proportional to the corresponding concentrations for small values of these concentrations. Different forms of suspension function $f(c)$ are used to match the experimental data [25, 26].

The initial-boundary value conditions for constant-concentration colloid injection into a clean bed are:

$$t = 0 : c = s = 0, \quad x = 0 : c = c^0 \tag{4}$$

where $c^0$ is particle concentration in the injected suspension.

Eq. (2) suggests gradual retention accumulation with time. During injection of one pore volume into a core with length $L$, $1/\alpha$ pore volumes are injected into the first core section with length $\alpha L$ ($\alpha <1$). So, for each $\alpha <1$, the total retained concentration in the first section of the core is higher than that in the overall core. Therefore, the retention profiles (RPs) given by solutions of system (1, 2) monotonically decrease. All available analytical models for deep bed filtration exhibit monotonic RPs [17, 27-35]. In contrast to this expected behaviour, numerous experimental studies of particle flows through porous media exhibit NRP. NRPs have been observed for pathogens [36-39], engineered NPs [2, 40, 41], and glass microspheres [38, 42].

NRP can be reproduced by the mathematical model that accounts for depth dependency of the filtration function $h(x,s)$, introduced by Bradford, et al. [43]. This dependency is considered to have power-law type; tuning the power yields high agreement between the laboratory and modelling data. Close match has been obtained for breakthrough curves (BTCs) and RPs. Presently, this model is widely used for interpretation of the laboratory data on colloidal transport [22, 44]. Yet, Goldberg, et al. [44] suggested that more physics justification for $x$-dependency of the filtration function is required.

The two-speed mathematical model to explain the NRP for monodispersed flow was developed by Yuan and Shapiro [45]. The slow particles are assumed to move along the grain surfaces into the stagnant zones, while the fast particles move in the bulk of the carrier fluid. Both fines attachment and detachment are considered. Seven empirical model parameters are tuned from the laboratory data. An excellent match of BTCs and RPs was achieved. Besides, the paper shows that two models with essentially non-equivalent physical assumptions may describe the same set of coreflood data. Knowledge of the micro-level filtration mechanisms is necessary to discriminate between the models.

Another two-velocity model to explain the NRPs was proposed by Bradford, et al. [46]. Neglecting retained-fines detachment into the bulk flux reduces the number of model coefficients to six. The authors highlight non-uniqueness of the tuning problem, where the same laboratory data can be reproduced by different sets of the model parameters. The model successfully matches numerous laboratory tests.

Two-speed effects are pronounced in heterogeneous media with large size of micro-heterogeneity, like layer-cake laminated or fractured-porous formations [26]. For low-dispersed heterogeneity, where distances between the high- and low-conductivity channels have an order of magnitude of the pore length, the Peclet's number has an order of magnitude of $10^{-3}$-$10^{-5}$, suggesting "instant" equalising of the suspension concentrations in the populations [47]. This yields effective single-population transport with the average "low" velocity [48, 49].

Currently, NRP is an important topic of significant and frequent debates [2, 5, 6, 12, 41, 42, 45, 50-55]. There is no consensus on the physics phenomena yielding NRPs.

The current paper explains non-monotonic retention behaviour by the colloid heterogeneity, where the attached particles with different properties (size, surface charge, form) occupy different areas on the rock surface. We discuss multicomponent colloidal flow in porous media without detachment accounting for pore accessibility and consequent flux reduction, slow particle motion and adsorption. The retention rate is a function of the vacant area for attachment to the rock (Langmuir's blocking) [56]. The micro-scale population-balance equations have first integrals along the characteristic lines; this allows for exact upscaling procedure, yielding the mass balance and kinetics equations for averaged concentrations. The upscaled 1D problem allows for exact solution. The analytical model exhibits NRP. The upscaled binary-colloid model successfully matches ten laboratory tests. Moreover, the dimension of the measured data array exceeds the number of tuned parameters, which validates the model. Tuning the model coefficients and using first integrals along the characteristic lines yield downscaling, i.e., determining the individual properties for each particle population. These results provide valuable insights for understanding the colloidal-nano transport phenomena exhibiting NRP.

The structure of the paper is as follows. Section 1 introduces the mathematical model for transport of uniform colloids and presents the physics explanations of NRPs. Section 2 provides a qualitative analysis of binary colloidal transport with attachment and predicts NRPs. Section 3 develops the exact upscaling procedure for multicomponent colloidal flows and derives the exact solution for upscaled model. Section 4 discusses the results, including type curves for BTC and RP and matching the experimental data. Section 5 concludes the paper.

## 2. Qualitative explanation of non-monotonic retention profiles

In this section we explain the occurrence of non-monotonic RP during colloidal injection with two distinct particle sizes. The large particles have a higher filtration coefficient (probability of particle capture per unit length of the trajectory) and occupy a higher proportion of area on the rock surface when attached.

Consider an early time ($t_1$) when the total area occupied by the retained particles is significantly lower than the initial vacant area on the rock surface. The particles do not compete for vacancies, so the two populations filter independently. The retention rate of large particles is higher than that of small particles at the inlet, as capture probability is higher for large particles. Consequently, the suspended and retained concentrations of large particles decline faster along the core. At some distance $\Delta$ from the core inlet, the suspended concentration of large particles is significantly lower than that of small particles. Therefore, as shown in Figs. 1a and 2a, near to the core inlet large-particle retention exceeds that for small particles, while small particle retention is predominant further in the core. Individual and total retention and suspension profiles monotonically decline across the core.

At later time ($t_2$), the area occupied by attached particles near the inlet reaches the order of magnitude of the initial vacant area (Fig. 1b). The particles are now competing for vacancies. As large particles that occupy higher area are retained predominantly near the inlet, vacant surface area increases rapidly along the core. Therefore, the retention rates of both populations is higher at positions further in the core than at the inlet, as there is higher vacant area.

At distance $\Delta$ from the core inlet, the suspended concentration of small particles is not significantly lower than its injected concentration; however, the amount of vacant surface area is significantly higher than at the inlet. So the retention rate of small particles at this point exceeds that at the inlet, and at some moment the retained concentration of small particles at the distance $\Delta$ exceeds that at the inlet. The small-particle retained concentration is zero at the suspended concentration front and increases from the inlet. Therefore, the small-particle RP has transitioned to non-monotonic (red curve in Fig. 2b). In Fig. 2b, its maximum is reached in the second core section.

As shown above, the suspended large-particle concentration always declines rapidly near to the inlet. As retention occurs in positions with low vacant area, the negative gradient of the large particle RP increases with time near to the inlet but the profile always monotonically decreases (green curve in Fig. 2c). The total RP monotonically decreases at initial stages but at some time ($t_3$) becomes non-monotonic (black curve). The maximum of total RP is reached in the third section of the core (Fig. 1c).

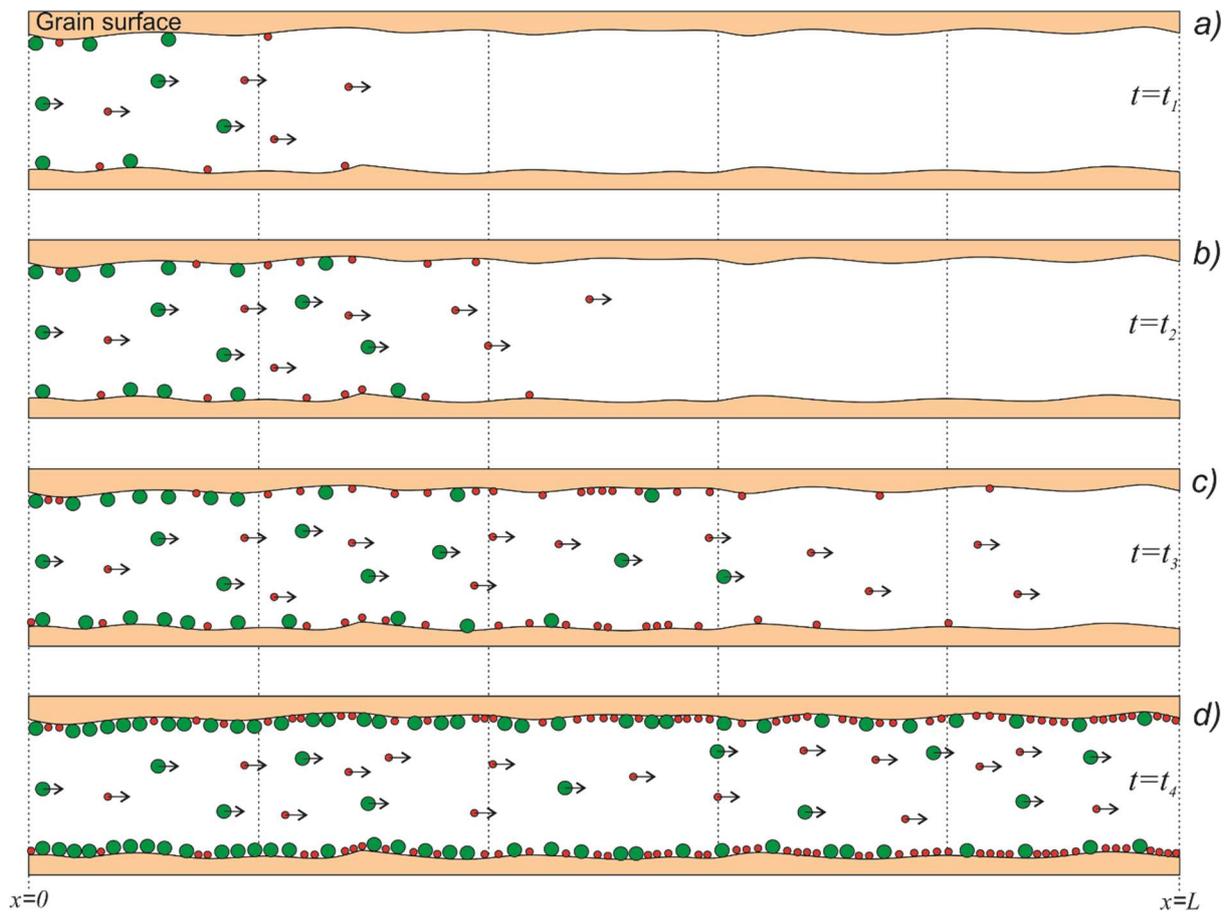

Fig. 1. Schematic showing retained particle distribution for four-stage evolution of retention profiles during simultaneous transport of two colloid size populations: a) large (green) and small (red) particles do not compete for vacancies, yielding declining profiles for both populations at an early stage of injection (moment $t_1$); b) at intermediate moment ($t_2$), large particles accumulate in the beginning of the core, while small particles travel further to accumulate within the core, resulting in non-monotonic profile for small particles; c) at the next intermediate moment ($t_3$), the total retention profile also becomes non-monotonic; d) at moment ($t_4$) all vacancies are occupied, with large and small particle quantities decreasing and increasing, respectively, with distance.

Further we refer to the maximum points as the positions where the retained concentrations are maximum. So far we have discussed the maximum points for small particle and total retained concentrations. Consider any point of the core ahead of the maximum point for small particles after the moment $t_3$. The vacant area at the maximum point is lower than at positions further into the core;

whereas, the suspended concentration is not significantly lower. So the retention rate for small particles is higher at positions ahead of the maximum point. Hence, the maximum point for small particles transitions through the core; i.e., the small particle RP monotonically increases over the core once the maximum point reaches the end of the core at the moment $t_4$ (Fig. 2d). By the same reasoning, the total RP also monotonically increases. As shown above, large particle retention never exceeds that at the inlet and monotonically decreases over the core. The three RPs asymptotically tend to their stabilised shapes as time tends to infinity, remaining monotonic.

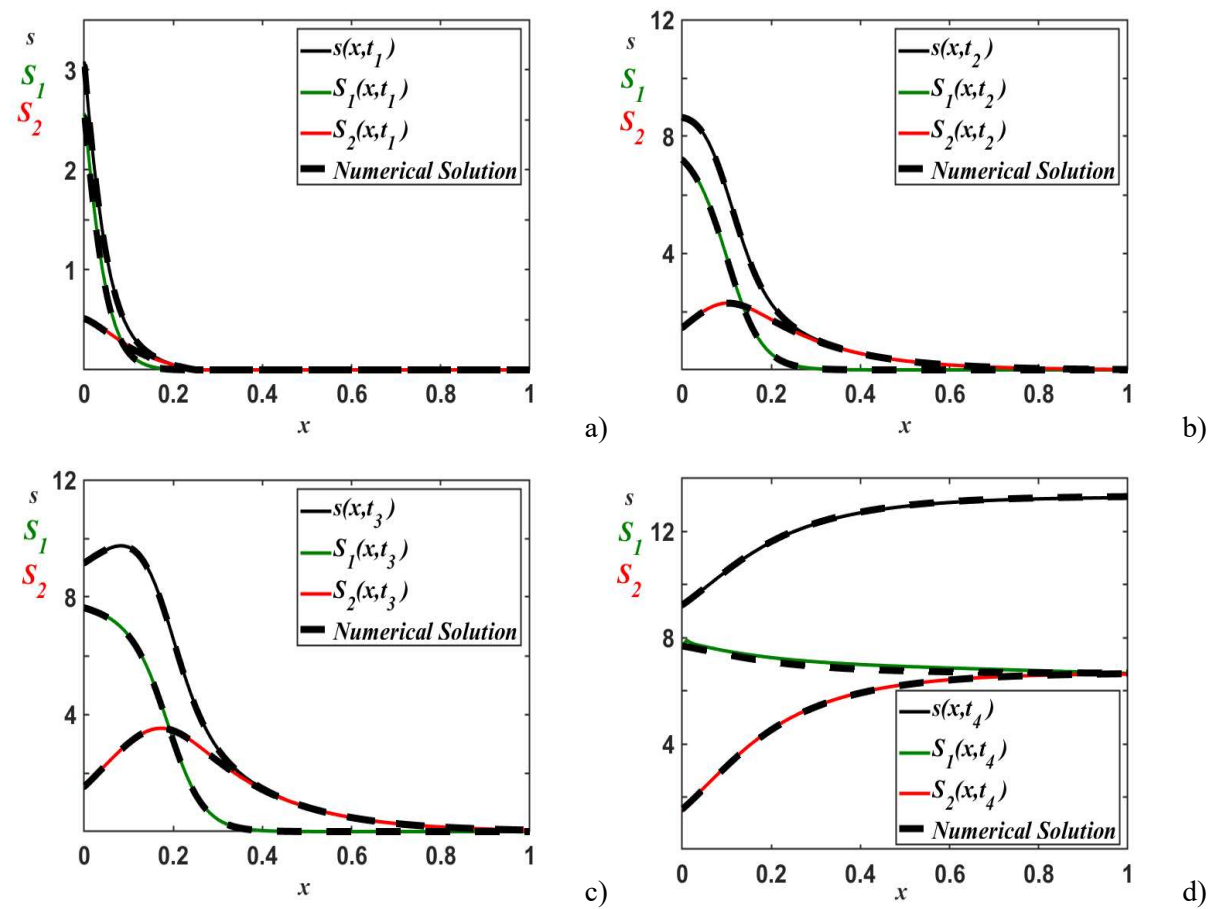

Fig. 2. Large-particle, small-particle, and total retention profiles (green, red and black curves, respectively) corresponding to moments $t_1=1$, $t_2=1.7$, $t_3=3$, and $t_4=50$ PVI: a) all profiles monotonically decrease; b) the small-particle profile becomes non-monotonic; c) the total profile becomes non-monotonic; d) stabilised large-particle profile monotonically decreases and stabilised small-particle and total profiles monotonically increase. Indexes 1 and 2 correspond to small and large particles, respectively.

### 3. Mathematical modelling

This section presents the model assumptions and system of governing equations (sections 3.1 and 3.2), develops an exact upscaling procedure (sections 3.3 and 3.4), derives the analytical model (section 3.5), develops the downscaling procedure (section 3.6), and verifies the analytical model by comparison with the numerical solution (section 3.7).

### 3.1 Model assumptions for multi-size colloidal transport

We consider a single-phase carrier fluid transporting colloidal suspension with multiple-size particles. The fluid and particles are incompressible. The volume of suspension is equal to the total of volumes of liquid and particles (Amagat's law) [57]. Those assumptions yield conservation of the overall suspension flux.

The particles can adsorb on the rock surface; the equilibrium adsorption with Henry's isotherm takes place [58]. The finite-size particles occupy $s_0$-th fraction of the porous space, called the accessibility factor [20, 43, 59-61]. The flux that carries particles through the accessible pore space is equal to $f_0$-th fraction of the overall flux of carrier fluid. The model (1, 2) accounts for low particle speed if compared with the carrier water, because the flux reduction factor $f_0$ contains the drift-delay factor [48, 49]. We assume that Henry's constant, accessibility and flux reduction factors are the same for all particles.

The filtration function is the same for all populations, while the suspension function depends on particle radius [26, 62]. So, different particle populations compete for the same attachment vacancies. Adsorption is assumed to be reversible, while the attachment is fully irreversible, i.e., no particle detachment occurs. Diffusion and dispersion fluxes for long cores are assumed to be negligibly small if compared with the advective fluxes [63, 64].

### 3.2 Governing equations for multiple populations

Basic equations for single-phase multi-component suspension-nano transport in porous media include mass balance of suspended, attached and adsorbed particles, and capture rate, for all populations [60, 62]:

$$\frac{\partial}{\partial t}\left(\phi s_0 \tilde{C}_k + \tilde{S}_k + \tilde{\Gamma}\tilde{C}_k\right) + U\frac{\partial}{\partial x}\left(f_0 \tilde{C}_k\right) = 0 \qquad (5)$$

$$\frac{\partial \tilde{S}_k}{\partial t} = h\left(\sum_{i=1}^{m}\tilde{B}_i\tilde{S}_i\right)F_k\left(\tilde{C}_k\right)f_0 U, \quad k=1,2...N \qquad (6)$$

where all parameters with tilde are dimensional, $\check{C}_k$ and $\check{S}_k$ are suspended and attached concentrations of $k$-th particle population, respectively, $\phi$ is the porosity, $\lambda_k$ is the filtration coefficient of each population, $\tilde{B}_k$ is the ratio between the area occupied by a single attached particle and the overall rock surface, and $\Gamma$ is Henry's sorption coefficient.

We use the following dimensionless parameters and variables:

$$x = \frac{\tilde{x}}{L}; t = \frac{U\tilde{t}}{\phi L}; c^0 = \sum_{i=1}^{m} C_i^0; C_k = \frac{\tilde{C}_k}{c^0}; S_k = \frac{\tilde{S}_k}{\phi c^0}; \lambda_k = \tilde{\lambda}_k L;$$
$$B_k = \phi c^0 \tilde{B}_k; A = \frac{\tilde{A}}{\phi c^0}; \Gamma = \frac{\tilde{\Gamma}}{\phi}; \ k = 1, 2, ..., N \quad (7)$$

where all parameters without tilde are dimensionless, and $L$ is the core length.

Substituting parameters (7) into system (5-6) we obtain the dimensionless governing system

$$\frac{\partial}{\partial t}(s_0 C_k + S_k + \Gamma C_k) + \frac{\partial}{\partial x}(f_0 C_k) = 0 \quad (8)$$

$$\frac{\partial S_k}{\partial t} = h\left(\sum_{i=1}^{m} B_i S_i\right) F_k(C_k) f_0, \ k = 1, 2...N \quad (9)$$

where $F_k(C_k)$ is the suspension function of $k$-th population.

The initial conditions and boundary conditions correspond to injection of colloids with constant composition into a clean bed:

$$t = 0: C_k = S_k = 0; \ x = 0: C_k = C_k^0 \quad (10)$$

System of equations (8, 9, 10) determines $2N$ unknowns $C_k$ and $S_k$.

### 3.3 Upscaling and system for total concentrations

In this section we derive the upscaled system for total suspended, retained and adsorbed concentrations. Substituting Eq. (9) into Eq. (8) yields

$$(s_0 + \Gamma)\frac{\partial C_k}{\partial t} + f_0 \frac{\partial C_k}{\partial x} = -h\left(\sum_{i=1}^{m} B_i S_i\right) F_k(C_k) f_0 \quad (11)$$

Characteristic form of first order partial differential equations (PDE) is a system of ordinary differential equations along characteristic lines in domain $f_0(s_0+\Gamma)^{-1}t > x$ [63, 64]

$$\frac{dt}{dx} = \frac{s_0 + \Gamma}{f_0}, \ \frac{dC_k}{dx} = -h\left(\sum_{i=1}^{m} B_i S_i\right) F_k(C_k) \quad (12)$$

Dividing variables in Eqs. (12) yields

$$\frac{dG_k(C_k)}{dx} = -h\left(\sum_{i=1}^{m} B_i S_i\right); \quad G_k(C_k) = \int_{C_k^0}^{C_k} \frac{du}{F_k(u)}, \quad k = 1, 2 \ldots N \tag{13}$$

The right hand sides of Eqs. (13) coincide for all $k=1,2\ldots N$. Functions $G_k$ are equal at $x=0$. Therefore, values $G_k(C_k(x,t))$ are equal for all $k$. In particular,

$$G_k(C_k) = G_1(C_1) \tag{14}$$

The total concentrations for suspended and retained particles and occupied area are determined as

$$c = \sum_{k=1}^{N} C_k, \quad s = \sum_{k=1}^{N} S_k, \quad b = \sum_{k=1}^{N} B_k S_k \tag{15}$$

Expressing individual concentrations $C_k$ from Eq. (14), accounting for monotonicity of functions $G_k$,

$$C_k = G_k^{-1}\left[G_1(C_1)\right], \tag{16}$$

substituting relationships (16) into Eq. (15) and expressing total concentration $c$ via $C_1$ yield

$$c = \sum_{k=1}^{N} G_k^{-1}\left[G_1(C_1)\right], \quad C_1 = g_1(c), \quad g_1 = \left[\sum_{k=1}^{N} G_k^{-1}(G_1)\right]^{-1} \tag{17}$$

Here symbol $()^{-1}$ corresponds to inverse function.

Substituting Eqs. (17) into Eqs. (16) leads to expressing each individual component $C_k$ versus total suspension concentration

$$C_k = G_k^{-1}\left[G_1(g_1(c))\right] = g_k(c) \tag{18}$$

So, all individual concentrations $C_k$ are functions of the total concentration $c$.

Summing Eqs. (15) into Eq. (8) yields total mass balance for all particle populations

$$\frac{\partial}{\partial t}(s_0 c + s + \Gamma c) + \frac{\partial(f_0 c)}{\partial x} = 0 \tag{19}$$

Multiplying each Eq. (9) by $B_k$, accounting for expressions (18), and summing the results lead to kinetics equation for site occupation:

$$\frac{\partial b}{\partial t} = h(b)d(c)f_0, \quad d(c) = \sum_{k=1}^{N} B_k F_k(g_k(c)) \tag{20}$$

where $d(c)$ is the occupation function.

Adding Eqs. (9) for $k=1,2\ldots N$ yields the kinetics equation for retention rate

$$\frac{\partial s}{\partial t} = h(b) f(c) f_0, \quad f(c) = \sum_{k=1}^{N} F_k(g_k(c)). \tag{21}$$

For $x < f_0(s_0+\Gamma)^{-1}t$, the solution is zero at both scales. System (19, 20, 21) describes 1D multicomponent colloidal-suspension transport. Initial condition $b=0$ adds to conditions (10) for injection into a clean bed.

**3.4   System for binary colloid transport**   Consider small injected concentrations where the suspension functions are linear, like in Eq. (3):

$$F_k(C_k) = \lambda_k C_k, \quad k = 1, 2 \tag{22}$$

Formulae (13, 16, 17) for $N=2$ become:

$$G_k(C_k) = \ln\left[\frac{C_k}{C_k^0}\right]^{1/\lambda_k}, \quad C_2 = C_2^0\left[\frac{C_1}{C_1^0}\right]^{\lambda_2/\lambda_1}, \quad c = C_1 + C_2^0\left[\frac{C_1}{C_1^0}\right]^{\lambda_2/\lambda_1}, \quad n = \frac{\lambda_1}{\lambda_2}; k=1,2 \tag{23}$$

yielding explicit expressions for suspension and occupation functions

$$f(c) = \lambda_1 g_1(c) + \lambda_2 C_2^0\left[\frac{g_1(c)}{C_1^0}\right]^{\lambda_2/\lambda_1} \tag{24}$$

$$d(c) = B_1\left\{\lambda_1 g_1(c) + \varepsilon \lambda_2 C_2^0\left[\frac{g_1(c)}{C_1^0}\right]^{\lambda_2/\lambda_1}\right\}, \quad \varepsilon = \frac{B_2}{B_1} \tag{25}$$

**3.5  Analytical model for multi-component colloidal transport**  Let us discuss the problem (10, 19, 20, 21) for Langmuir's filtration function

$$h(b) = 1 - b \tag{26}$$

The solution is obtained by method of characteristics,[63, 64] similarly to the solution of simplified colloidal transport [59, 62]. Introduce Lagrangian coordinate $\tau$:

$$\tau = t - \left(\frac{s_0 + \Gamma}{f_0}\right)x \tag{27}$$

System (19, 20, 21) in coordinates $(x, \tau)$ becomes

$$\frac{\partial c}{\partial x} = -(1-b)f(c) \tag{28}$$

$$\frac{\partial s}{\partial \tau} = (1-b)f(c)f_0 \qquad (29)$$

$$\frac{\partial b}{\partial \tau} = (1-b)d(c)f_0 \qquad (30)$$

Ahead of the concentration front that moves with velocity $f_0(s_0+\Gamma)^{-1}$, i.e. for $\tau < 0$, the initial conditions hold.

Suspension concentration behind the front $c^-$, where $b=0$ at $\tau=0$, is obtained from the relationship along the characteristic line for Eqs. (28):

$$\int_{c^-\left(x,-\frac{s_0+\Gamma}{f_0}x\right)}^{1} \frac{dy}{f(y)} = x \qquad (31)$$

The expression for suspended concentration $c(x,\tau)$ is obtained by expressing occupied vacancy concentration $b(x,\tau)$ from Eqs. (28, 30), integrating in $x$, and is given by the transcendental equation:

$$\int_{c^-\left(x,-\frac{s_0+\Gamma}{f_0}x\right)}^{c} \frac{f_0^{-1}dy}{f(y)w(y)} = \tau; \quad w(c) = \int_{c}^{1} \frac{d(y)}{f(y)} dy \qquad (32)$$

Taking $x$-derivative of both sides of Eq. (32), expressing gradient of suspended concentration and substituting the result into Eq. (28), we express occupied concentration $b(x,\tau)$ via suspended concentration:

$$b(x,\tau) = 1 - \frac{w(c)}{w(c^-)} \qquad (33)$$

The retained concentration $s(x,\tau)$ is also expressed via suspended concentration:

$$s(x,\tau) = \frac{c(x,\tau) - c^-(x)}{w(c^-)} \qquad (34)$$

The stabilised RP under the injected colloid flow in the rock with fully occupied vacancies for $\tau \to \infty$ is:

$$s(x,\infty) = \frac{1 - c^-(x)}{w(c^-(x))} \qquad (35)$$

### 3.6 Downscaling: determining individual population properties from total-concentration data

Expressing individual concentrations via total concentration in Eqs. (9) facilitates downscaling:

$$C_k(x,t) = C_k^0 \left[ \frac{g_1(c(x,t))}{C_1^0} \right]^{\lambda_k/\lambda_1}, \quad S_k(x,t) = f_0 \int_0^t (1-b(x,y)) F_k(C_k(x,y)) dy, \tag{36}$$

i.e. the individual concentrations can be obtained from matching the large-scale model using the laboratory data on both BTCs and RPs.

**3.7 Comparison between the analytical and numerical models** We validate the analytical model (25-29) by comparison with the numerical solution using the Shampine [65], [66] The two-step Lax–Friedrichs finite-difference method [65, 66] is inbuilt into MATLAB computer software (http://faculty.smu.edu/shampine/current.html). We divide the spatial domain [0,1] into 1000 evenly spaced intervals, with the time step selected to satisfy the Courant–Friedrichs–Lewy stability condition, $\Delta t = 0.9\Delta x$.

Figure 2 presents the comparison-study results for the following parameters: $B_2=0.025$, $B_1=0.125$ ($\varepsilon=0.2$), $C_2^0=0.5$, $\lambda_2=5$, and $\lambda_1=25$ ($n=5$). Close agreement between the exact and numerical solutions is observed for RP at four typical moments.

# 4 Results and discussion

In this section we present the type curves for suspended and occupation functions, based on the upscaling procedure, and investigate the forms of BTCs and RPs, based on the analytical model (section 4.1). Section 4.2 presents matching of the laboratory data and tuning the model coefficients. Sensitivity study of BTCs and RPs with respect to flux reduction coefficient and adsorption constant are given in section 4.3. Section 4.4 presents the summary of upscaling and exact integration and discusses their extension to more complex flow systems.

**4.1 Analysis of 1D analytical modelling** Analytical modelling presented in Fig. 2 exhibits monotonically decreasing RP at the initial stage of injection, appearance of a peak in the RP for small-particles during the second stage, occurrence of non-monotonic total RP at the third stage, and monotonically increasing profiles at stabilisation. So, the analytical modelling fully reproduces the qualitative behaviour of binary-colloid flow predicted by the qualitative analysis (Fig. 1).

Fig. 3 shows the effects of the ratios $n = \lambda_1/\lambda_2$ and $\varepsilon = B_2/B_1$ on suspension and occupation functions, BTCs and RPs.

Increase of ratio $n$ under fixed filtration coefficient for small particles yield an increase in $\lambda_1$, so the suspension function increases. Green curves in Fig. 3a are located above the corresponding blue curves. Increase in injected concentration of small particles $C_2^0$ and consequent decrease in $C_1^0$ causes the suspension function to decrease. Dashed curves in Fig. 3a are located above the corresponding continuous curves. For small total concentration ($c\rightarrow 0$), suspension function degenerates into linear function (3).

Fig. 3b shows that if the ratio of occupied areas $\varepsilon$ increases under fixed occupied area by small particles $B_2$, the area $B_1$ decreases, area $b$ is occupied at lower rate, filtration function $h$ increases; as it follows from kinetics (21) the occupation function $d(c)$ decreases. Total suspended concentration also decreases, i.e. the BTC moves down (Fig. 3c).

For $\varepsilon=n$ the occupation function $d(c)$ becomes linear. This corresponds to the same area being occupied by each population per unit of time. For all cases where $\varepsilon\neq n$, $d(c)$ is nonlinear. The larger the difference between $B_1$ and $B_2$, the larger the curvature of the occupation function. The $d(c)$-curves are convex for $\varepsilon<n$ and concave for $\varepsilon>n$ (Fig. 3b).

The red curves in Fig. 3d are RPs corresponding to the case where large particles attach with higher intensity and occupy larger area than small particles ($\lambda_1>\lambda_2$ and $B_1>B_2$). The profile is monotonic and convex at the beginning of injection ($t_1$). It remains monotonic but at the moment $t_2$ the profile becomes concave near to the inlet. The profile then becomes non-monotonic at the moment $t_3$ and stabilises with concave monotonically increasing shape at the moment $t_4$.

Black curves in Fig. 3d correspond to the case where large particles attach with lower intensity ($\lambda_1<\lambda_2$) and occupy higher area than small particles ($B_1>B_2$). The RPs always keep the monotonically decreasing form.

The effect of the filtration coefficient ratio $n=\lambda_1/\lambda_2$ on the RP are shown in Fig. 4. The first three stages shown in Fig. 1 are reproduced. The higher the filtration ratio $n$, the faster that non-monotonicity appears and the more pronounced is the non-monotonic behaviour. For fixed filtration ratio, increase in the filtration coefficient for small particles yields increase in the retained concentration.

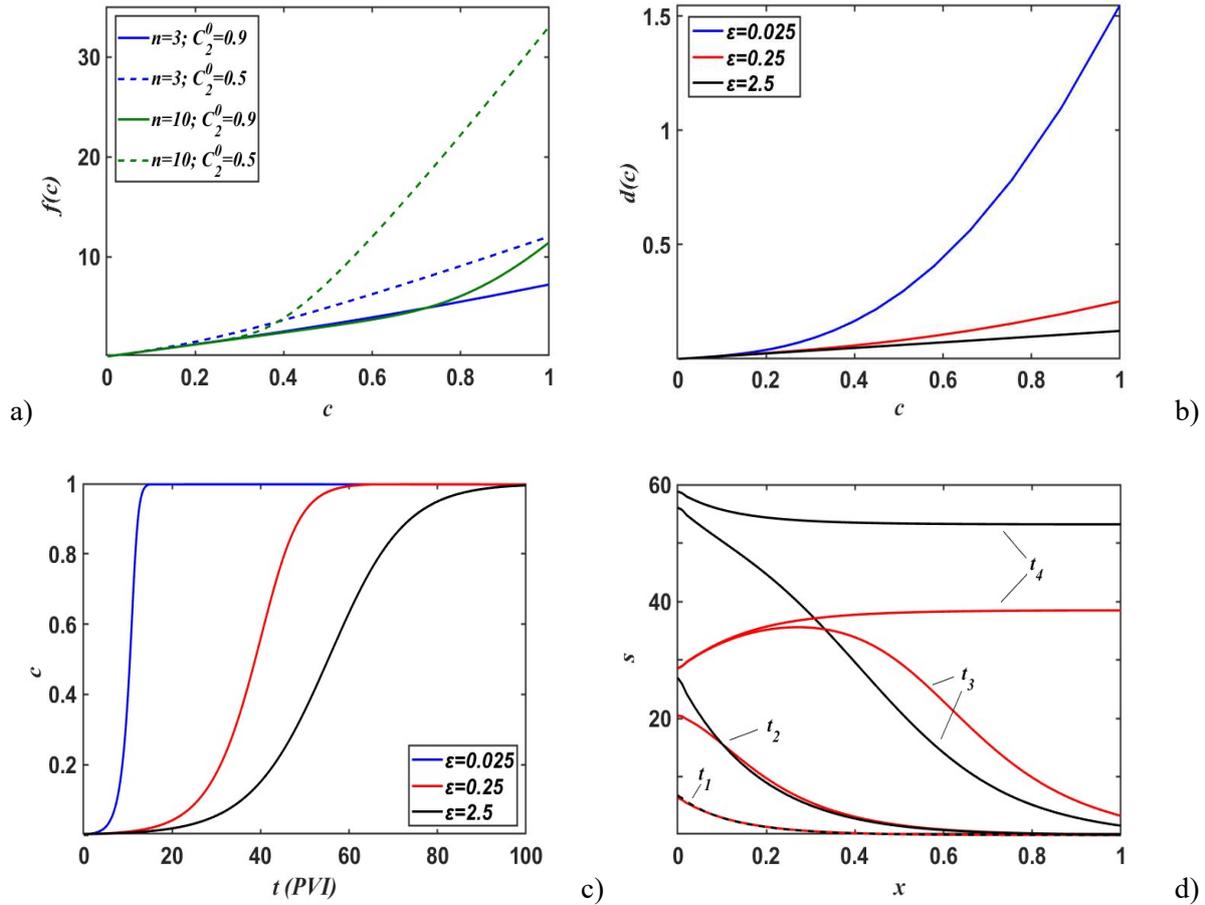

Fig. 3. Type curves for the model functions and solutions: a) suspension function for different ratios between filtration coefficients $\lambda_2$ and $\lambda_1$, $n = 3$ and 10, and injection concentrations $C_2^0 = 0.5$ and 0.9; b) occupation function for different ratios of occupied areas $B_2$ and $B_1$ ($\varepsilon = 0.025$, 0.25, and 2.5); c) effect of $\varepsilon$ on the breakthrough concentration profile; d) retention profiles for $\varepsilon = 0.25$ (red curves) and $\varepsilon = 2.5$ (black curves) at the moments $t_1 = 1$, $t_2 = 5$, $t_3 = 25$ and $t_4 = 100$ PVI. Fixed parameters used in the modelling for b, c and d) are $B_2 = 0.02$, $C_2^0 = 0.9$, $\lambda_2 = 6$; $n = 3$.

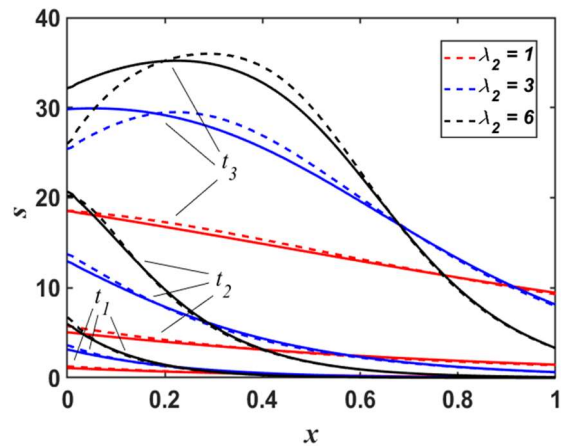

Fig. 4. Effect of $n = \lambda_1/\lambda_2$ on the retention profile: $n = 2$ (continuous curves) and $n = 4$ (dashed curves) for $\lambda_2 = 1$, 3, and 6; at 3 moments $t_1 = 1$, $t_2 = 5$, and $t_3 = 25$. Fixed parameters used in the modelling are $B_2 = 0.02$, $C_2^0 = 0.9$, $\varepsilon = 0.25$.

## 4.2 Matching the experimental data

Fig. 5 shows the laboratory data on injection of silver NPs into soil specimen with length 10 cm, diameter 1.1 cm and with three salinities 1, 5, and 10 mM for $KNO_3$ (Figs. 5a and 5b) and 0.1, 0.5 and 1 mM for $Ca(NO_3)_2$ (Figs. 5c and 5d) [41]. The model parameters - filtrations coefficients and occupied areas for both populations and the injected concentration – are estimated by least-square minimisation of the deviation between the laboratory and modelling breakthrough curves and retention profiles (RPs). The matching is performed by the nonlinear least squares method [67] implemented as an internal optimisation procedure within Matlab. Simultaneous matching of BTCs and RPs is performed by tuning five model coefficients $\lambda_2$, $n$, $C_2^0$, $B_2$ and $\varepsilon$. We consider no adsorption mechanism ($\Gamma=0$), no flux reduction $f_0=1$, and an accessibility factor of $s_0=1$. The values of five tuned model coefficients are presented in Table 1. The BTCs have at least three free parameters and the number of degrees of freedom for RPs is at least four. All matched curves are located insides the areas bounded by the error bars. The tenth and eleventh columns in Table 1 reveal coefficient-of-determination values close to one for the bulk of the data. Close match of seven-parametric array of experimental data by five model parameters validates the model.

Matching of silver NPs injection into quartz sand column also exhibit close agreement between the experimental and analytical modelling data (rows 8 and 9 of Table 1) [50]. Rows 10 and 11 correspond to the tests on de-ionised water injection with $TiO_2$ NPs into white sand column [2].

The model parameters determined from 20 PVI's history (red curves in Figs. 5a and 5b) allow revealing long-term total and individual breakthrough histories until stabilisation (50 PVI) as shown in Fig. 6a. The corresponding total and individual RPs near stabilisation are revealed in Figs. 6b and 6c at moment $t_4$. Moment $t_3$ in Fig. 6c reveals the individual RPs at termination of the experiment. Individual and total retention profiles at earlier moments $t_1$ and $t_2$ are also shown in Figs 6b and 6c. The behaviour of the individual and total retention profiles match that qualitatively predicted in Fig. 1.

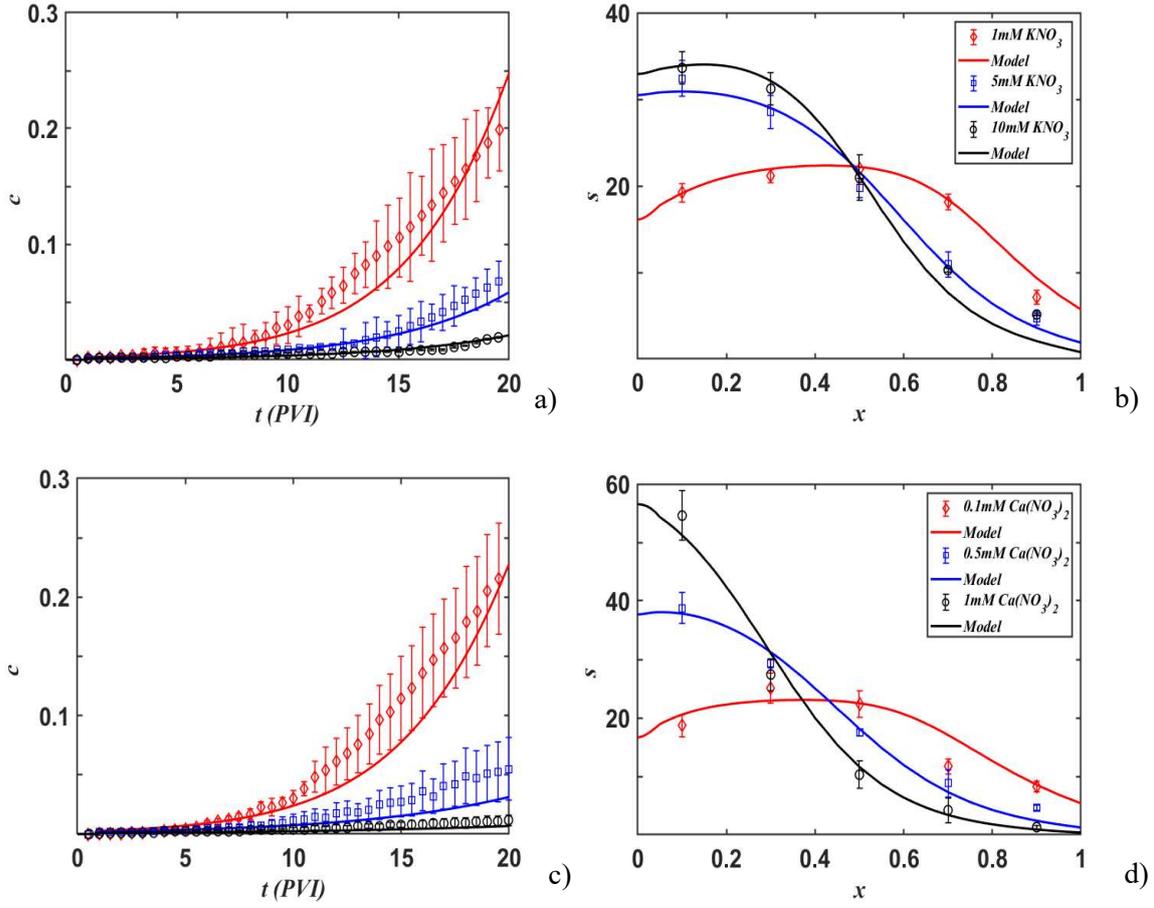

Fig. 5. Matching of the laboratory data by the model for different salts and salinities: a) and c) breakthrough concentration profiles for $KNO_3$ and $Ca(NO_3)_2$, respectively; and the corresponding retention profiles b) and d) at $t = 20$ PVI.

### 4.3 Sensitivity study

Let us discuss how the flux reduction factor $f_0$ and the adsorption constant $\Gamma$ affect the BTCs and RPs during the corefloods.

Fig. 7 shows BTCs and RPs for both populations and the total. The injected concentrations for both populations are the same, so the suspended concentrations stabilise at the same value (Fig. 7a). For the case modelled, stabilisation occurs at $t= 20$ PVI. The total RP and individual RPs are shown at the moments $t_1 = 1$, $t_2 = 5$, $t_3 = 10$ and $t_4 = 20$ PVI at Figs. 7b and 7c, respectively. Those four moments represent four typical stages for the deep bed filtration of binary colloid (Figs. 1).

Figs. 7d, e, and f correspond to slow particle filtration $f_0$=0.01 [48]. Change of time variable $t \rightarrow f_0 t$ transforms system (19, 20, 21) to the same system but with $f_0$=1. If compared with the case, the evolution in system (19, 20, 21) is delayed $1/f_0$ times. BTCs in Figs. 7a and 7d coincide, if the time axis in Fig.

7d is expanded $1/f_0$ times. RPs in Figs. 7b and 7e coincide if the four PRs in the case $f_0=0.01$ are taken at the instances $t_k/f_0$, k=1…4. The same corresponds to individual RPs in Figs. 7c and 7f.

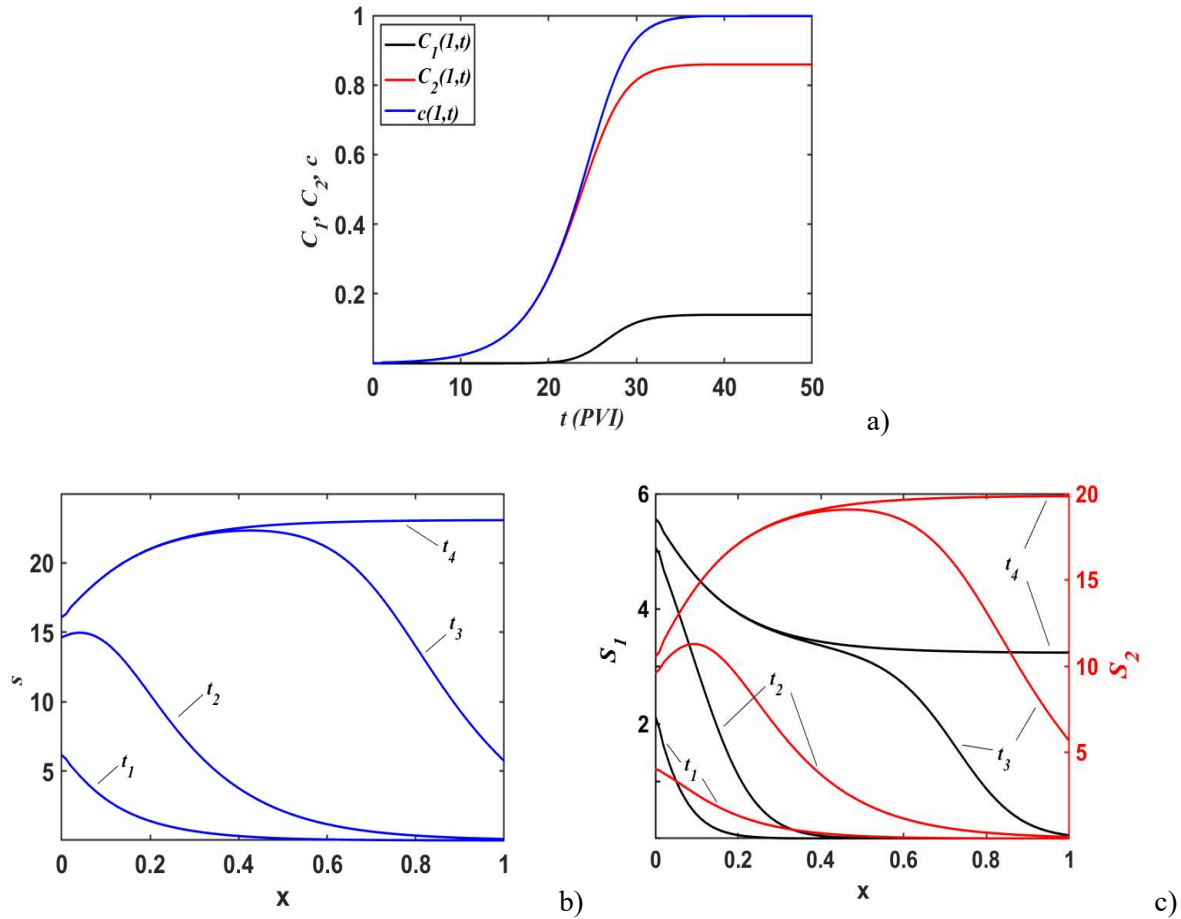

Fig. 6. Revealing micro- and macro-scale behaviour from the matched laboratory data: a) breakthrough concentration profiles for each component and the total colloid; b) revealed earlier and later retention profiles from limited-time matching; c) retention profiles for each component at moments $t_1 = 1$, $t_2 = 5$, $t_3 = 20$, and $t_4 = 50$ PVI for 1mM $KNO_3$ (Fig. 5b).

The effects of Henry's adsorption constant on the colloidal-nano transport is shown in Fig. 8. Figs. 8a, b, and c compare the adsorption-free case with the case $\Gamma = 1$. BTCs with sorption repeat the forms of the sorption-free case with delay 1 PVI (Figs. 8a). Pink, green, and blue curves in Fig. 8b correspond to retained, adsorbed and total concentrations, respectively. The total concentration repeats type curves for retained concentrations as obtained in the four typical moments. This result is important, because there is no way to distinguish between adsorbed and retained particles in porous media during laboratory corefloods, they cannot be measured separately, and what profile is measured by CT is the total RP [19, 25, 45].

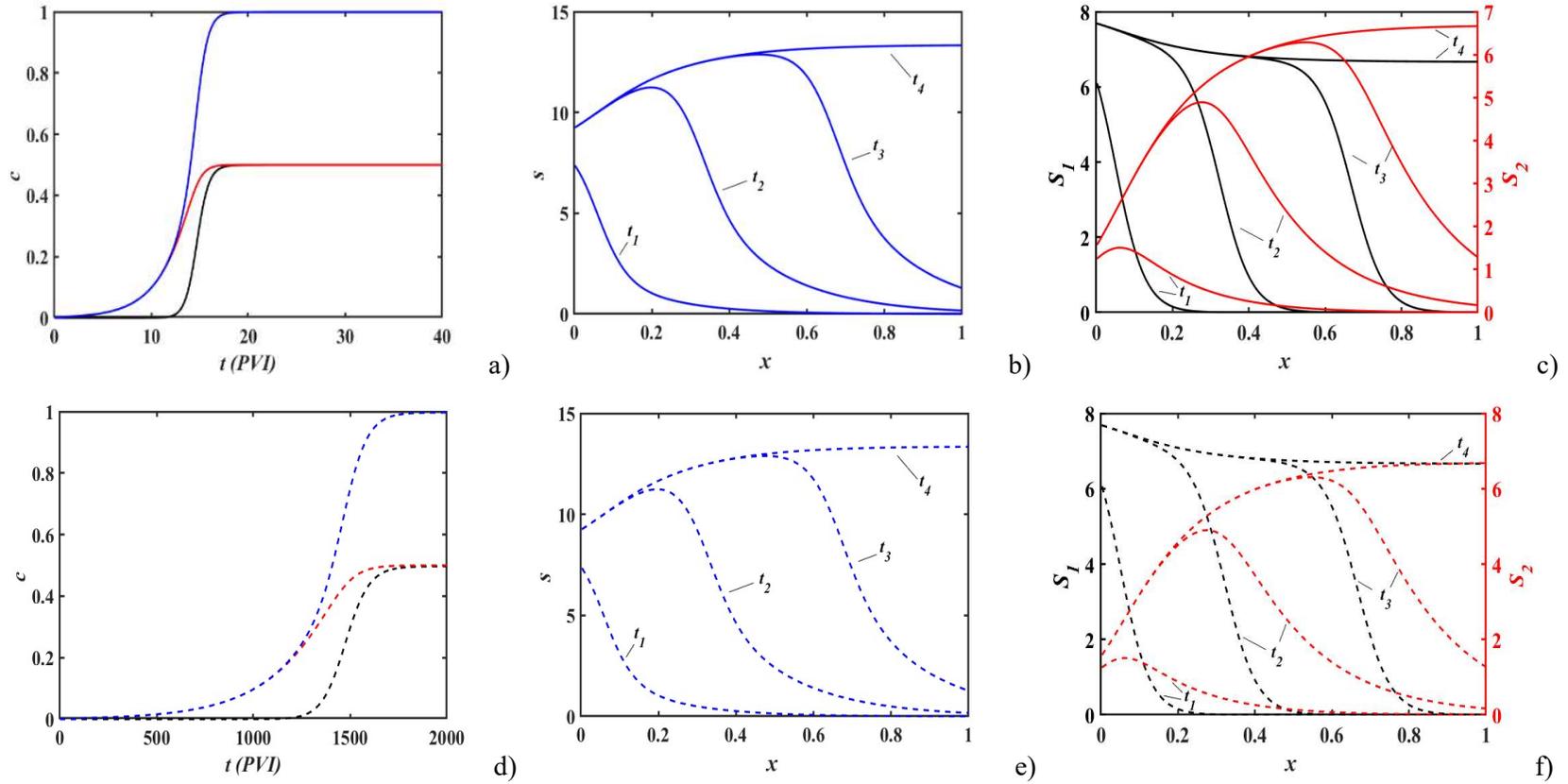

Fig. 7. Comparison between BTCs and RPs for small-particle population (red), large-particle population (black) and total colloid (blue) for different $f_0$. For a), b) and c) $f_0$=1 at the moments $t_1$= 1, $t_2$= 5, $t_3$= 10 and $t_4$=20 PVI; for d), e) and f) $f_0$=0.01 at the moments $t_1$= 100, $t_2$= 500, $t_3$= 1000 and $t_4$=2000 PVI. Parameters used for analytical model: $B_2$=0.025; $\varepsilon$=0.2; $C_2^0$=0.5; $\lambda_2$=5; n=5; $\Gamma$=0; $s_0$=0.8.

Red and black curves in Fig. 8c correspond to small and large retained particles, respectively. At the presence of adsorption, the individual RPs also follow the four-stage scenarios of sorption-free case. At the case of sorption, the RPs of each population repeat the sorption-free profiles with delay. So, equilibrium sorption results in delay in colloidal transport.

Figs. 8d, e, and f compare the cases $\Gamma=0$ and $\Gamma=4$. The effects are the same as in the previous case, but the delay is larger (4 PVI).

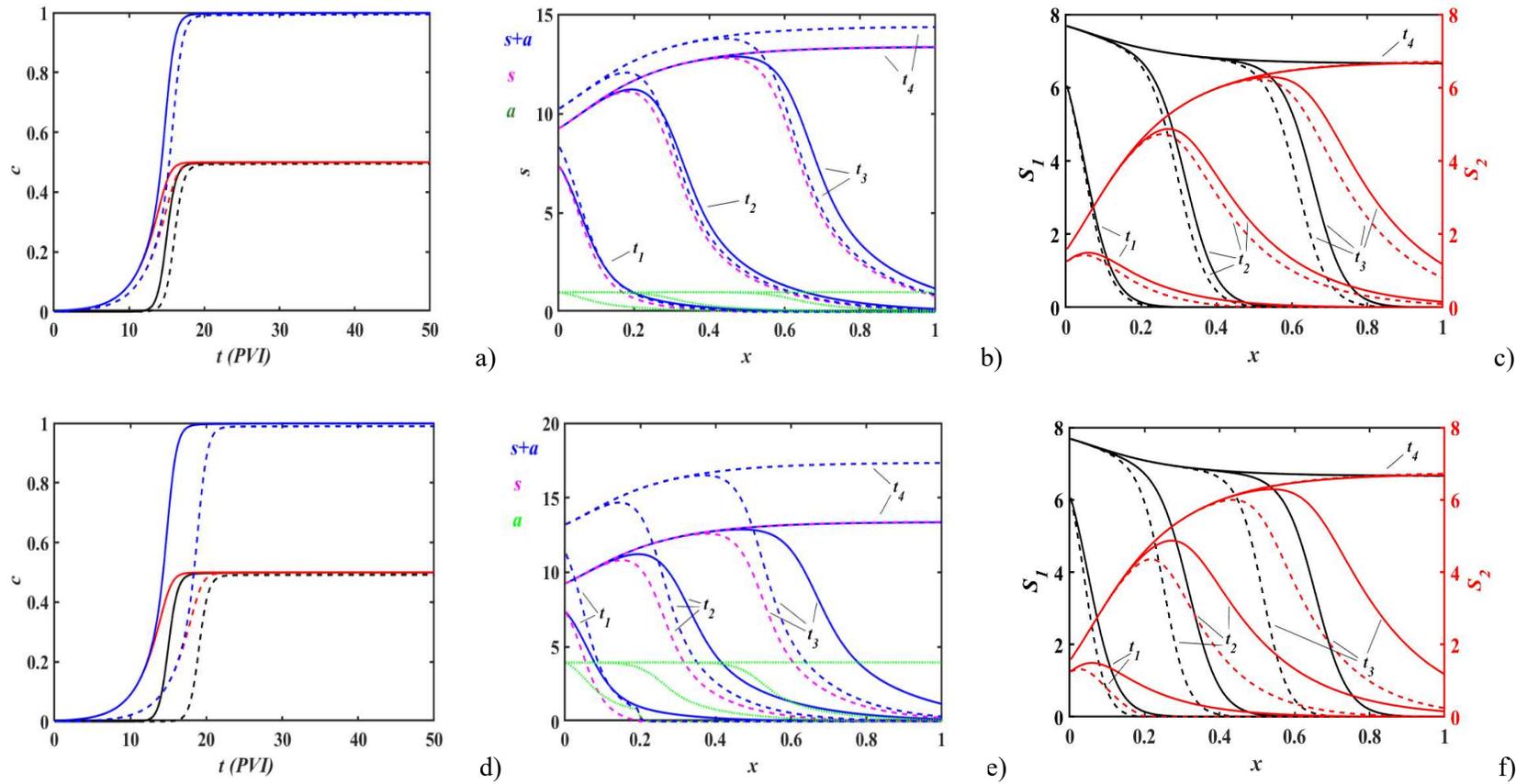

Fig. 8. Comparison of BTCs for macro- and micro-scale and RPs for macro- and micro-scale for different Henry's adsorption coefficient $\Gamma$. For a), b) and c) $\Gamma=0$ (continuous curves) and $\Gamma=1$ (dashed curves); for d), e) and f) $\Gamma=0$ (continuous curves) and $\Gamma=4$ (dashed curves). Parameters used for the analytical modelling are $B_2=0.025$; $\varepsilon=0.2$; $C_2^0=0.5$; $\lambda_2=5$; $n=5$; $s_0=1$; $f_0=1$ at the moments $t_1=1$, $t_2=5$, $t_3=10$ and $t_4=20$ PVI

## 4.4 Upscaling of more complex systems. Generalisations of the upscaling procedure. Applications.

NRPs have been observed in numerous laboratory studies and explained by depth-dependency of the filtration function, and by two-speed structure of the colloidal flux [40, 43, 45, 46]. The corresponding models account for fines attachment and detachment. However, our results suggest that non-monotonic variation of the attached particle concentration along the core can be reproduced by the simpler model that accounts for attachment only.

This is the upscaled (averaged) model for colloidal transport of size-distributed particles, where each particle population flow fulfils the traditional deep-bed filtration equations. We discuss colloids with stochastically distributed particle properties that affect attachment (size, shape, surface charge) and account for attachment, adsorption, accessibility, and slow particle motion. The resulting upper-scale model, along with mass-balance and attachment-rate kinetics, contains an equation for site-occupation kinetics. Despite the upscaled 3×3 hyperbolic system is non-linear, 1D transport problem allows for exact solution. The upscaled analytical model facilitates tuning the model coefficients from the laboratory coreflood data and allows for the laboratory-based predictions for 3D field-scale behaviour using stream-line modelling. This determines the impact of the upscaling technique and the analytical model.

The upscaling procedure (11-21) permits for several extensions. Consider travelling waves in system of PDEs (19, 20, 21), where the solutions depend on one variable $x-Dt$, where $D$ is travelling-wav velocity. It results in system of ODE, coinciding with Eqs. (28, 29, 30). The following derivations (31-34) are valid for the system of ODE yielding the upscaled model.

Consider suspension injection with varying concentrations $C_k^0(t)$. It yields time-dependency of the lower limit in integral (13). Consequently, in right hand sides of dependencies (14, 15, 16) appear variable $t$, resulting in time-dependencies of suspension and occupation functions in the upscaling system (19, 20, 21).

Zero boundary conditions and non-zero initial conditions (10) correspond to migration of natural suspended reservoir fines [25, 62]. The upscaling procedure (11-21) also yields the averaged system (19, 20, 21). For non-uniform initial fines distribution $C_k^0(x)$, suspension and occupation functions in the upscaling system (19, 20, 21) are $x$-dependent.

Eqs. (5, 6) for colloidal-nano transport in porous media are similar to the reactive flow models, which often do not allow for upscaling [68, 69]. In this case, the upper scale system depends on the micro-scale behaviour, and the hybrid models with simultaneous simulations at different scales are applied [70, 71].

Table 1. Tuned model coefficients as obtained from laboratory data

| Ref. | $KNO_3$ (mM) | $Ca(NO_3)_2$ (mM) | Velocity (cm/min) | $\phi$ | $\lambda_2$ | $\lambda_1$ | $C_2^0$ | $B_2$ | $B_1$ | $R_c^2$ | $R_s^2$ |
|---|---|---|---|---|---|---|---|---|---|---|---|
| [41] | 1.0 | - | 0.26 | 0.4 | 5.90 | 19.18 | 0.860 | 0.0305 | 0.122 | 0.9493 | 0.9625 |
|  | 5.0 | - | 0.26 | 0.4 | 6.30 | 11.53 | 0.930 | 0.0242 | 0.103 | 0.9438 | 0.9933 |
|  | 10.0 | - | 0.26 | 0.4 | 7.15 | 11.44 | 0.940 | 0.0200 | 0.118 | 0.8669 | 0.9641 |
|  | - | 0.1 | 0.26 | 0.4 | 5.80 | 29 | 0.895 | 0.0350 | 0.103 | 0.9456 | 0.8175 |
|  | - | 0.5 | 0.26 | 0.4 | 6.15 | 18.45 | 0.980 | 0.0225 | 0.075 | 0.6326 | 0.9848 |
|  | - | 1.0 | 0.26 | 0.4 | 7.00 | 9.8 | 0.990 | 0.0140 | 0.14 | 0.4710 | 0.9860 |
| [50] | 2.5 | - | 0.70 | 0.4 | 1.50 | 15 | 0.670 | 0.1000 | 0.345 | 0.8494 | 0.9378 |
|  | 5.0 | - | 0.70 | 0.4 | 2.90 | 23.2 | 0.580 | 0.0300 | 0.25 | 0.8411 | 0.9964 |
| [2] | - | - | 3.00 | 0.38 | 7.00 | 63 | 0.975 | 0.0035 | 0.041 | - | 0.9010 |
|  | - | - | 6.80 | 0.38 | 6.00 | 33 | 0.975 | 0.0045 | 0.075 | - | 0.8653 |

## 5 Conclusions

Analysis of NRPs by the means of upscaling the multi-component colloidal flow system that accounts for adsorption and pore accessibility and 1D analytical model yield the following conclusions.

Under the assumption that the particles that occupy larger area on the rock surface are attached with larger probability, the analytical model of macro-scale equations exhibits NRPs. This assumption is more realistic than the opposite one, so non-monotonic behaviour is a rule rather than exception.

Qualitative analysis allows distinguishing four typical periods for of colloidal transport of a binary mixture: a) initial stage with monotonically declining RPs; b) appearance of the concentration "hill" of small particles at the inlet and its movement along the flux in the core; c) appearance of the "hill" for overall particle concentration at the inlet and its movement along the flux; d) occupation of the core by ascending sides of both hills and monotonically increasing stabilised profiles for small-particle and overall concentrations.

This scenario is fully supported by concentration-wave behaviour of the analytical model.

The model for colloidal transport of binary mixture exhibits close simultaneous match of laboratory data for BTCs and RPs. Moreover, the data array with 7 degrees of freedom was matched by the 5-parametric model, which validates the model.

If smaller particles are captured faster ($B_1<B_2$, $\lambda_1> \lambda_2$), retention profiles remain monotonically decreasing during the overall injection period.

The exact upscaling includes expressing each individual suspended concentration versus the averaged value. It allows for complete downscaling, i.e. determining the individual population concentrations $C_k(x,t)$ and $S_k(x,t)$ from their upscaled values $c(x,t)$ and $s(x,t)$.

The stabilised profiles correspond to trivial steady-state solution of the upscaled system: $b=1$, and $c=1$. However, stabilised retention profile $s(x,t)$ with $t\rightarrow\infty$ cannot be found from the upscaled system. It is found using the Riemann invariants.